\documentclass[conference]{IEEEtran}
\IEEEoverridecommandlockouts
\usepackage{cite}
\usepackage{amsmath,amssymb,amsfonts}
\usepackage{algorithmic}
\usepackage{graphicx}
\usepackage{textcomp}
\usepackage[utf8]{inputenc}
\usepackage[hyphens]{url}
\usepackage{balance}
\usepackage[numbers]{natbib}
\usepackage{multirow}
\usepackage{colortbl}
\usepackage{tabularx}
\usepackage{subcaption}
\usepackage{float}
\usepackage{xcolor}
\usepackage{booktabs}
\usepackage{adjustbox}
\usepackage{subcaption}
\usepackage{color, colortbl}
\definecolor{Gray}{gray}{0.95}
\usepackage{enumitem}
\usepackage[bottom]{footmisc}

\def\BibTeX{{\rm B\kern-.05em{\sc i\kern-.025em b}\kern-.08em
    T\kern-.1667em\lower.7ex\hbox{E}\kern-.125emX}}
\usepackage{fancyhdr}
\usepackage{kantlipsum}
\fancypagestyle{plain}{
\fancyhf{}
\fancyhead[C]{Conference on \LaTeX} 

}
\usepackage{eso-pic}
\begin{document}

\title{Characterizing (Un)moderated Textual Data in Social Systems}

\author{\IEEEauthorblockN{Lucas Lima\IEEEauthorrefmark{1}, Julio C. S. Reis\IEEEauthorrefmark{1}\IEEEauthorrefmark{2}, Philipe Melo\IEEEauthorrefmark{1}, Fabrício Murai\IEEEauthorrefmark{1}, Fabrício Benevenuto\IEEEauthorrefmark{1}}
\IEEEauthorblockA{\IEEEauthorrefmark{1}Universidade Federal de Minas Gerais (UFMG), Brazil, \IEEEauthorrefmark{2}Universidade FUMEC, Brazil}
\{lucaslima, julio.reis, philipe, murai, fabricio\}@dcc.ufmg.br
}




\maketitle


\begin{abstract}

Despite the valuable social interactions that online media promote, these systems provide space for speech that would be potentially detrimental to different groups of people. The moderation of content imposed by many social media has motivated the emergence of a new social system for free speech named Gab, which lacks moderation of content. This article characterizes and compares moderated textual data from Twitter with a set of unmoderated data from Gab. In particular, we analyze distinguishing characteristics of moderated and unmoderated content in terms of linguistic features, evaluate hate speech and its different forms in both environments. Our work shows that unmoderated content presents different psycholinguistic features, more negative sentiment and higher toxicity. Our findings support that unmoderated environments may have proportionally more online hate speech. We hope our analysis and findings contribute to the debate about hate speech and benefit systems aiming at deploying hate speech detection approaches.
\end{abstract}


\section{Introduction}
\label{sec:1_Introduction}

The Web has changed the way our society communicates, giving rise to social platforms where users can share different types of content and freely express themselves through posts containing personal opinions. Unfortunately, with the popularization of this new flavor of communication, toxic behaviors enacted by some users have been gaining prominence through online harassment and hate speech. These platforms have become the stage for numerous cases of online hate speech, a type of discourse that aims at attacking a person or a group on the basis of race, religion, ethnic origin, sexual orientation, disability, or gender~\cite{johnson2019hidden}.

Recently, to prevent the proliferation of toxic content, most online social networks prohibited hate speech in their user policies and enforced this rule by deleting posts and banning users who violate it. Particularly, Twitter, Facebook, and Google (YouTube) have largely increased removals of hate speech content~\cite{reu-hate}
by making available their hate policies so users can actively report content that might violate their policies. Reddit also deleted some communities related to fat-shaming and hate against immigrants~\cite{Chandrasekharan:2017:YCS:3171581.3134666}. This scenario has motivated the emergence of a new social network system, called Gab. In essence, Gab is very similar to Twitter, but barely moderates any of the content shared by its users. According to Gab guidelines, the website promotes freedom of expression and states that ``\textit{the only valid form of censorship is an individual’s own choice to opt-out}''. They, however, do not allow illegal activity, spam, or form of illegal pornography, promotion of violence and terrorism.

Despite existing recent efforts that attempt to understand the content shared in Gab~\cite{lima2018inside, zannettou2018gab}, there is still a need for an understanding regarding the amount and forms of hate speech in an unmoderated system such as Gab. 
In this article, we provide a diagnostic of hate in the unmoderated content from Gab, by categorizing the different forms of hate speech in that system and comparing it with Twitter, a proxy for a moderated system. Specifically, we identify textual characteristics of a set of unmoderated (or barely moderated) data, in this work represented by Gab, and compare them with characteristics of moderated data, here represented by Twitter. Our study is based on the analysis of $7,794,990$ Gab posts and $9,118,006$ tweets from August 2016 to August 2017. At a high level, our analysis is centered around the following research questions:
\noindent \textit{\textbf{ Research Question 1:}} \textit{What are the distinguishing characteristics of moderated content in Twitter and unmoderated content in Gab in terms of linguistic features, sentiment, and toxicity?}

\noindent \textit{\textbf{Research Question 2:}} \textit{What are the most common types of hate in an unmoderated and moderated environment?}

To answer our first research question we quantify the linguistic differences across moderated and unmoderated social systems by applying established language processing techniques. Our analysis shows that content in Gab and Twitter have different linguistics patterns, with higher toxicity and a more negative overall sentiment score in the unmoderated Gab content. 
Additionally, we find that, in general, Gab has more hate  posts than Twitter. 
We show that Gender and Class types of hate are more frequent on Twitter, whereas Disability, Ethnicity, Sexual Orientation, Religion, and Nationality types tend to appear proportionality more in Gab.

These findings highlight the importance of creating moderation policies as an effort to fight online hate speech in social systems, and also point out possible points for improvement in the design of content policies for social media systems. Additionally, our findings suggest that the unmoderated content found in Gab might be an appropriate data source for the development of learning approaches to detect hate speech. Thus, as a final contribution, we make our hate-labeled Gab posts available for the research community\footnote{\url{https://github.com/lhenriquecl/unmoderated-hate-dataset}} 
can foster the development of future hate speech detection systems.

\section{Related Work}
\label{sec:2_RelatedWork}

\subsection{Social Media Content Moderation}

The ongoing discussion on content moderation in an online environment led Internet companies to reflect about the undesirable attention their sites can attract and the consequences of it\footnote{\url{https://www.nytimes.com/2010/07/19/technology/19screen.html}}, thus the first steps towards regulating and moderating content have already been taken. There is a lot of debate on how social media platforms moderate their own content, and how their moderation policies are shaped. 
Twitter and YouTube, for instance, make available their hate policies\footnote{https://help.twitter.com/en/rules-and-policies/hateful-conduct-policy}\textsuperscript{,}\footnote{https://support.google.com/youtube/answer/2801939?hl=en} so users can actively report content that might violate their policies. 
Our work contributes to this discussion since we quantify the differences between moderated and unmoderated text data as an effort to shed light on the importance of creating different methodologies and policies to outline the boundaries of hate in social media.

Besides, some studies have focused on the understanding of systems that lack moderation of content. Finkelstein et al.~\cite{finkelstein2018quantitative} focus on making an extensive analysis of antisemitism in 4chan's and Gab. Their results provide a quantitative data-driven framework for understanding this form of offensive content. Zannettou et. al \cite{zannettou2018gab} and Lima et. al \cite{lima2018inside} present the first characterization studies on Gab, analyzing network structure, users, and posts. Our effort is complementary to these works as we compare the textual data shared from an unmoderated system like Gab with a moderated one as Twitter and deeply investigate the types of hate in both social systems.

\subsection{Online Manifestations of Hate Speech}

A vast number of studies were conducted to provide a better understanding of hate speech on the Internet. Silva et al.\ and Mondal et al.\ \cite{silva2016analyzing, mondal2017measurement} provide a deeper understanding of the hateful messages exchanged in social networks, studying who are the most common targets of hate speech in these systems. Salminen~\mbox{
\cite{salminen2018anatomy} }\hspace{0pt}
create a taxonomy and use it to investigate how different features and algorithms can influence the results of the hatefulness classification of text using several machine learning methods. 
Chandrasekharan et al.\ \cite{chandrasekharan2017you} characterize two banned communities from \textit{Reddit}, one about fat shaming and the other related to hate against immigrants in the US, and proposed a lexicon for hate speech detection. Our effort is complementary to these studies, as we quantify the amount of hate shared in a moderated and in an unmoderated environment and highlight the different types of hate exchanged in these social systems, elucidating the importance of content moderation to fight hate on the Internet. 

Several other efforts have attempted to provide detection approaches for hate speech~\cite{bartlett2014anti,gitari2015lexicon,agarwal2015using,warner2012detecting}. 
As a final contribution, we make the hate-labeled Gab dataset available to the research community. We hope our efforts help other hate speech studies on the creation of better methods for identifying hate on social media. 
\section{Datasets and Methods }
\label{sec:3_BackgroundInformation}

\subsection{Datasets}

Our Gab dataset comprises posts from users crawled following a Breadth-First Search (BFS) scheme on the graph of followers and friends. We used as seeds users who authored posts listed by categories on the Gab main page. We implemented a distributed and parallel crawler that ran in August 2017. In total, our Gab dataset comprises $12,829,976$ posts, obtained from $171,920$ users (the estimated number of users in August 2017 was $225$ thousand~\cite{mashable-gab}). 

The Twitter dataset contains English posts randomly selected from the Twitter 1\% Streaming API.  For consolidating our dataset and keep data consistency, we consider only random tweets published in the same period as Gab posts, which gives us also $12,829,976$ tweets. After preprocessing and removing duplicated posts in both datasets, we have a total of \textbf{7,794,990 Gab posts} and \textbf{9,118,006~tweets}. These are the final sets of posts for each media that are going to be further analyzed in this work.

\subsection{Language Processing Methods}

Next, we describe the methods used in this work to perform language characterization.

\subsubsection{Linguistic Analysis} One of our goals is to understand the distinguishing linguistic characteristics of posts on Gab and Twitter and contrast them. Thus, we use the $2015$ version of the Linguistic Inquiry and Word Count (LIWC) \cite{tausczik2010psychological} to extract and analyze the distribution of psycholinguistic elements posts of both media.  LIWC is a psycholinguistic lexicon system that categorizes words into psychologically meaningful groups all of which form the set of LIWC attributes, and has been widely used for several different tasks~\cite{Ribeiro2016, reis2019supervised, correa-2015-anonymityShades, resendewebsci19}. 

\subsubsection{Sentiment Analysis} 
We perform sentiment analysis on Gab and Twitter posts as a complementary effort to characterize the differences between moderated and unmoderated accounts. 
We use an established opinion mining method to measure sentiment score on our messages: the SentiStrength~\cite{thelwall2013heart}, which has shown to be an effective tool for sentiment analysis in social media posts \cite{abbasi2014benchmarking}. 
We apply the standard English version of it to quantify positive \textit{P}~$\in\{+1,\ldots,+5\}$ and negative \textit{N}~$\in\{-5,\ldots,-1\}$ sentiments in each post, as well as their overall sentiment score, which is given by the difference between \textit{P} and absolute \textit{N} values for a post.

\subsubsection{Toxicity Analysis} We measure the toxicity of posts with the Perspective API, created by Jigsaw and Google’s  Counter Abuse Technology team, also as a complementary analysis to elucidate the difference between moderated and unmoderated content. This score measures how ``\textit{toxic}'' a message can be perceived by a user.
Toxic messages are defined as \textit{a rude, disrespectful, or unreasonable comment that is likely to make you leave a discussion}. The value does not represent a degree of ``toxic severity'' of a particular message, but instead the probability that someone perceives that message as toxic. Scores range from 0 to 1, where scores closer to 1 indicate that posts are likely to be perceived as toxic. 


\subsection{Assessing Hate Speech}
\label{subsec:Hate}

ElSherief et al.~\cite{elsherief2018peer} present a semi-automated classification approach for the analysis of directed explicit hate speech which relies on keyword-based methods and on the Perspective API. The authors validate their methodology by incorporating human judgment using Crowdflower, concluding that their final hate speech dataset is reliable and has minimal noise. The method to detect hate implemented in this work is inspired by the referred work and it is similar to it with minor changes.

\begin{figure}[t]
    \center {\includegraphics[scale=0.4]{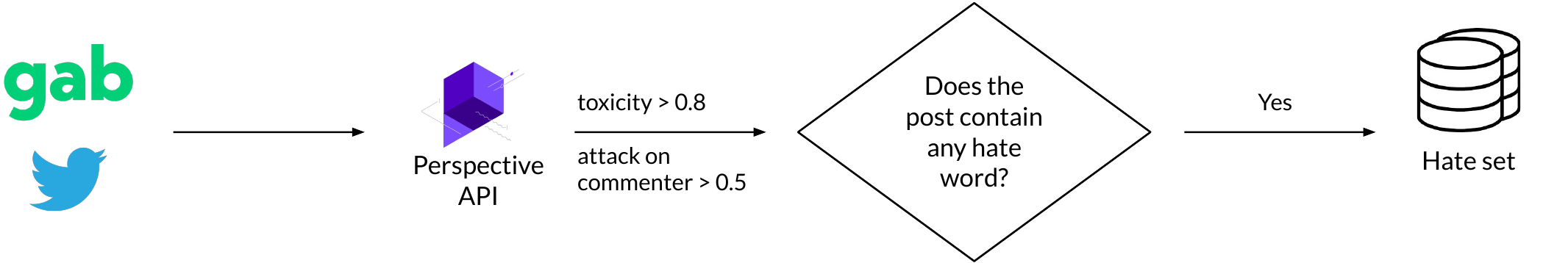}}
    \caption{Flowchart representing the method for identifying hate posts from Gab and Twitter posts.}
    \label{fig:methodology}
    \vspace{-3mm}
\end{figure}

Figure~\ref{fig:methodology} illustrates the method for identifying hate posts implemented in this work. First, both datasets go through a pipeline where the initial step is to query the Perspective API using each post as input. Besides the toxicity score, we gather the \textit{attack on commenter} score of posts, which measures direct and personal offense or injury to another user participating in the discussion. Next, we filter posts which have toxicity score higher than 0.8 and attack on commenter higher than 0.5 (these thresholds were defined by ElSherief et al.~\cite{elsherief2018peer} so as to yield a high quality dataset). Finally, we check whether these filtered posts contain at least one hate word, and, if so, we assume these are hate posts.

\begin{table}[t]
\centering
\caption{Lexicon of categorized hate terms.}
\resizebox{\columnwidth}{!}{
\begin{tabular}{|l|l|}
\hline
\textbf{Ethnicity}                                                    & \begin{tabular}[c]{@{}l@{}}bamboo coon, boojie, camel fu**er, house ni**er,\\ moon cricket, ni**er, plastic paddy, raghead, \\ sideways pu**y, spic, trailer park trash, trailer trash, \\ wetback, whi**er, white ni**er, white trash, wi**er, zionazi\end{tabular} \\ \hline
\textbf{Class}                                                        & \begin{tabular}[c]{@{}l@{}}bitter clinger, boojie, redneck, rube, trailer park trash,\\ trailer trash, white trash, yo**o\end{tabular}                                                                                                                               \\ \hline
\textbf{Disability}                                                   & retard, retarded                                                                                                                                                                                                                                                     \\ \hline
\textbf{Nationality}                                                  & \begin{tabular}[c]{@{}l@{}}bamboo coon, camel fu**er, chinaman, limey, plastic paddy, \\ sideways pu**y, soup taker, surrender monkey, whi**er, \\ white ni**er, wi**er, zionazi\end{tabular}                                                                        \\ \hline
\textbf{Religion}                                                     & camel fu**er, muzzie, soup taker, zionazi                                                                                                                                                                                                                            \\ \hline
\textbf{Gender}                                                       & bint, cu*t, d*ke, t*at                                                                                                                                                                                                                                               \\ \hline
\textbf{\begin{tabular}[c]{@{}l@{}}Sexual\\ Orientation\end{tabular}} & d*ke, fa**ot                                                                                                                                                                                                                                                         \\ \hline
\end{tabular}}
\label{table:hateWords}
\vspace{-6mm}
\end{table}

The list of hate words is also obtained from the study of ElSherief et al.~\cite{elsherief2018peer}. 
Table~\ref{table:hateWords} shows the terms\footnote{Wherever present, the `*' has been inserted by us, in order to lessen the impact that the offensive terms may inflict on some people, and was not part of the original word or text} from the categorized lexicon which are currently in Hatebase and are associated with at least one type of hate. Following the aforementioned methodology, \textbf{9,554} (0.12\%) Gab posts and \textbf{2,392} (0.03\%) tweets are labeled as \textbf{hate} and are going to be further explored on our hate speech analysis.

\section{RQ1: Distinguishing Characteristics of Moderated and Unmoderated Content}
\label{sec:4_LinguisticDifferences}

\subsection{Linguistic Features}

\begin{figure}[t]
    \center {\includegraphics[scale=0.25]{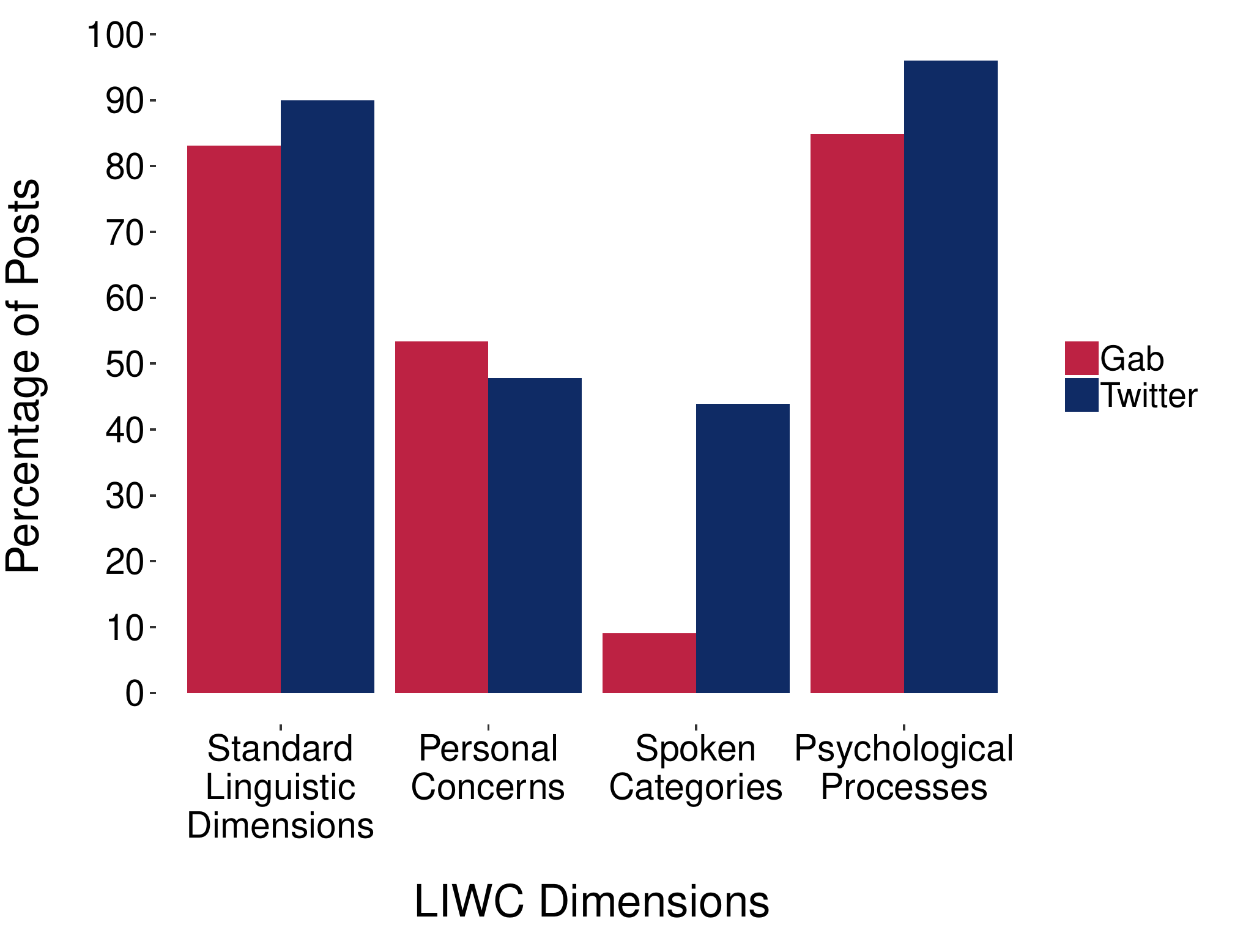}}
    \caption{Percentage of Gab and Twitter posts which contain at least one word or token per LIWC dimension.}
    \label{fig:percentagePostsLIWC}
    \vspace{-6mm}
\end{figure}

We analyze linguistic differences of moderated and unmoderated content by computing the distributions values for each LIWC attribute in both sets of posts. We aggregate these attributes into four distinct dimensions (\textit{Standard Linguistic Dimensions}, \textit{Personal Concerns}, \textit{Spoken Categories}, and \textit{Psychological Processes}) following Pappas et al.~\cite{pappas2016stateology} who made this arrangement available\footnote{\url{https://lit.eecs.umich.edu/geoliwc/liwc_dictionary.html}}. We start by investigating the volume of posts from both social media containing words in each LIWC dimension, as shown in Figure~\ref{fig:percentagePostsLIWC}.

 More than 80\% of Gab and Twitter posts contain at least one term of either the Standard Linguistic Dimensions or the Psychological Processes dimensions. Nearly 50\% of posts from both social media contain words of Personal Concerns. Interestingly, for the Spoken Categories, 43.9\% of Twitter posts contain at least one word of this dimension, whereas only 9.1\% of Gab posts contain at least one of the referred words. This difference might be due to the characteristics of the audience and posts of the Gab network, which has a strong political bias where users tend to share a larger number of news and politics related posts~\cite{lima2018inside}, whereas Twitter has traits of behavior that may encourage informal communication~\cite{zhao2009and}. We compare the distributions of each LIWC attribute for both Gab and Twitter by running the Kolmogorov-Smirnov (KS) test \cite{massey1951kolmogorov}.
 We find significant statistical difference (p-value $< 0.05$) for all the distributions, indicating that moderated and unmoderated posts have different psycholinguistic features. 

\subsection{Sentiment Analysis and Toxicity}

\begin{figure}[t]
\begin{subfigure}{0.23\textwidth}
\includegraphics[width=1\linewidth]{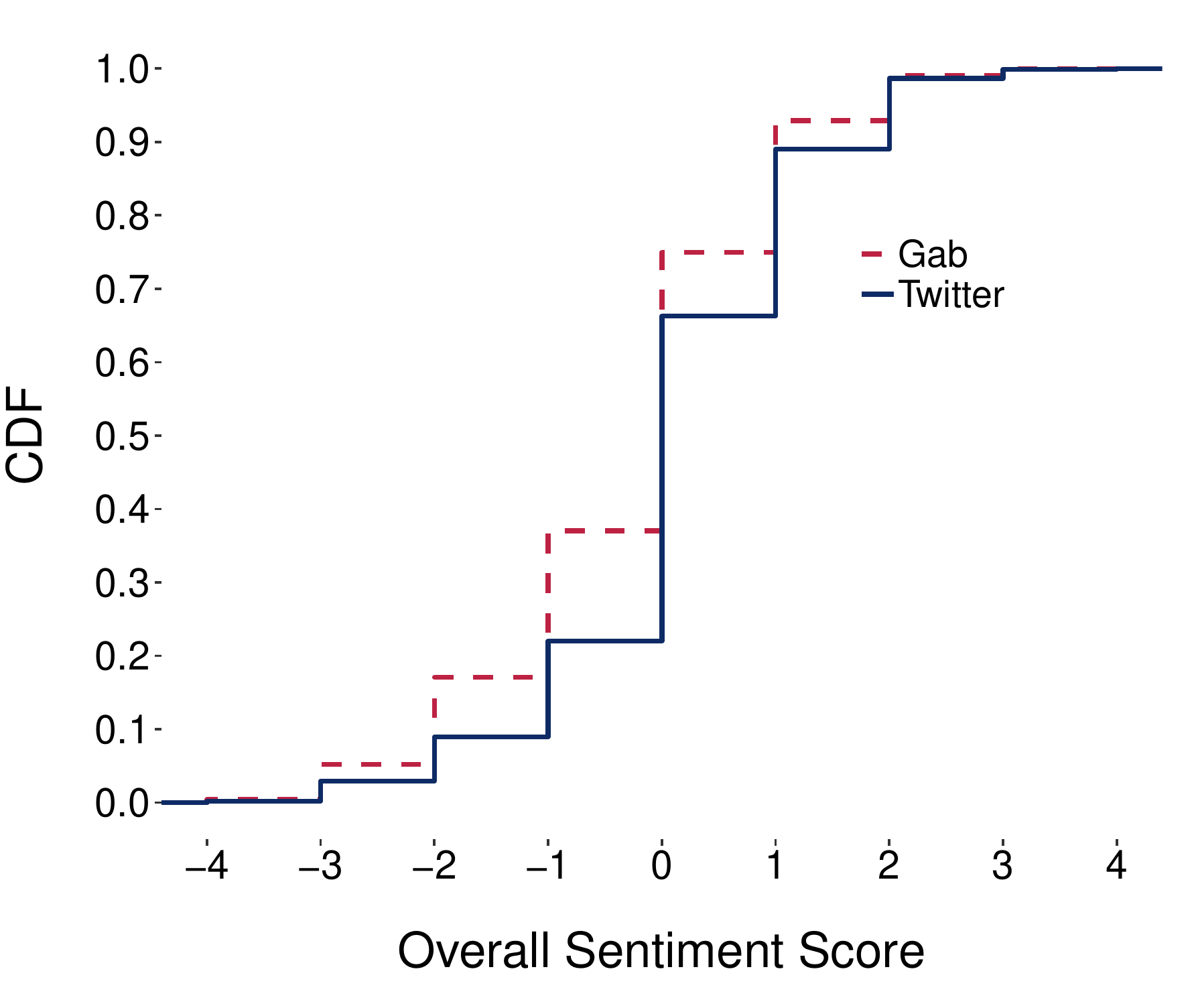}
\caption{}
\label{fig:CDFSentiment}
\end{subfigure}
\begin{subfigure}{0.23\textwidth}
\includegraphics[width=1\linewidth]{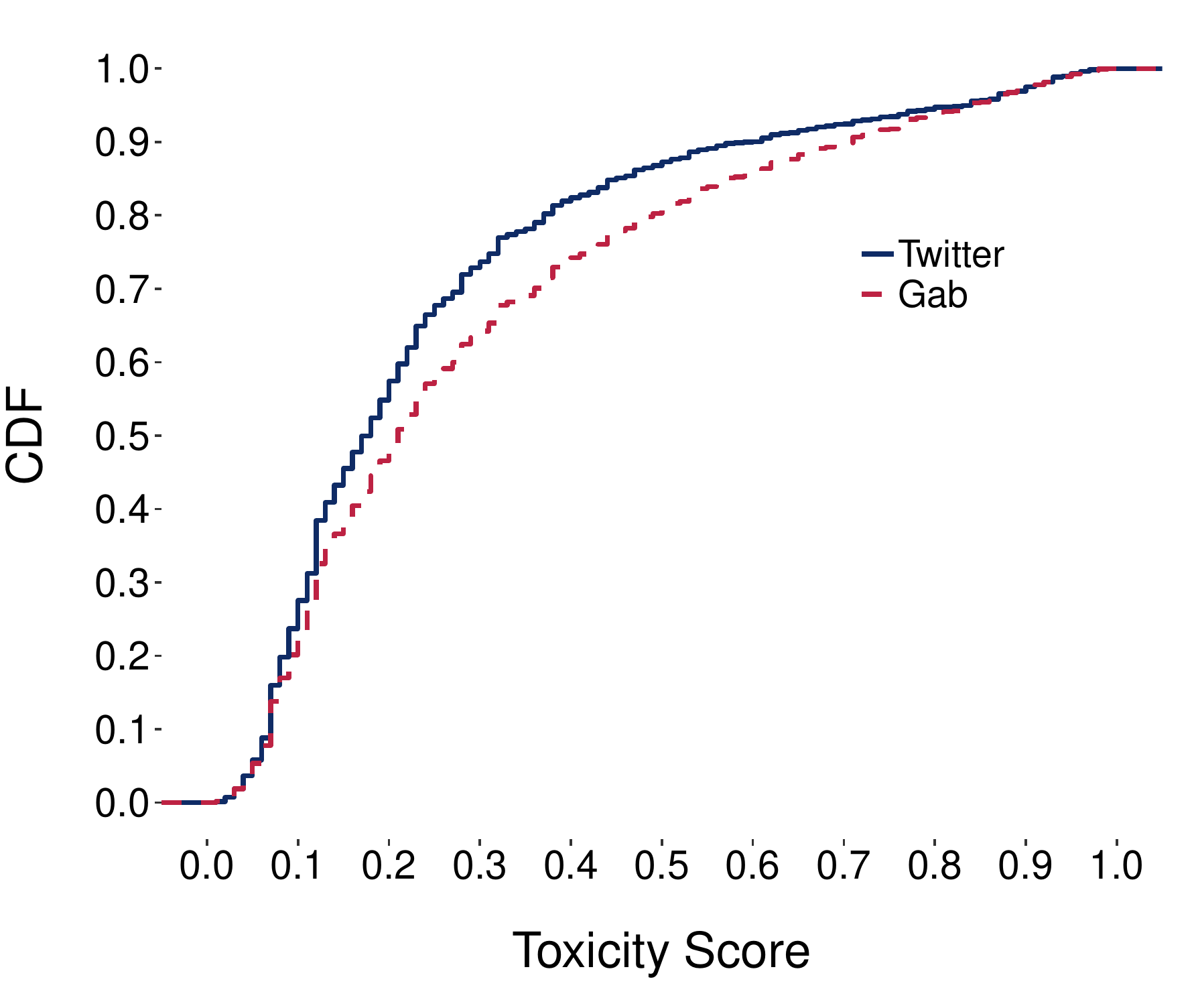}
\caption{}
\label{fig:CDFToxicity}
\end{subfigure}
\caption{CDF for (a) overall sentiment and (b) toxicity scores.}
\label{fig:CDFSentimentToxicity}
\vspace{-3mm}
\end{figure}

Next, we analyze the differences of sentiment and toxicity for moderated and unmoderated content. Figure~\ref{fig:CDFSentimentToxicity} shows the Cumulative Distribution Function (CDF) for the (a) overall sentiment score, calculated as the difference between the positive and the absolute value of negative scores given by SentiStrength ($P-|N|$), and (b) toxicity score for both Gab and Twitter. First, we compare these distributions using the KS test. For each metric, we find a significant statistical difference between the distributions, i.e., posts from moderated and unmoderated social media are statistically different in terms of sentiment and toxicity scores.

We notice that unmoderated publications tend to be more negative and to have higher toxicity than moderated ones. For the overall sentiment scores, Figure~\ref{fig:CDFSentiment} shows that nearly 37\% of unmoderated posts have negative overall sentiment, i.e., have an absolute negative score greater than a positive score, whereas only 22\% of moderated posts have negative overall sentiment. Furthermore, Figure~\ref{fig:CDFToxicity} shows that 19.4\% of Gab posts have toxicity scores higher than 0.5, whereas this percentage is 13\% for Twitter posts, indicating that unmoderated posts tend to be perceived as toxic more often than moderated posts. The fraction of posts that have toxicity above 0.8 is higher in Gab than on Twitter, as there are 6.5\% of such posts on Gab and 5.5\% of tweets in all posts. These results suggest that Gab is a more toxic social network than Twitter. One probable explanation for this is the lack of moderation on Gab, allowing harmful speech in this network to fester unchecked.
\section{RQ2: Different Types of Hate Across Moderated and Unmoderated Environments}
\label{sec:5_HateDifferentMedia}

\subsection{Manual Validation}

\begin{table}[t]
\centering
\caption{Manual evaluation. Each triple shows the number of posts agreed as hate, without agreement and agreed as non-hate, respectively.}
\begin{tabular}{@{}lll@{}}
\toprule
                                          & \multicolumn{1}{c}{\textbf{Gab}} & \multicolumn{1}{c}{\textbf{Twitter}} \\ \midrule
\textbf{Labeled by framework as Hate}     & (93, 4, 3)                       & (90, 6, 4)                           \\
\textbf{Labeled by framework as Non-Hate} & (5, 4, 91)                       & (0, 3, 97)                           \\ \bottomrule
\end{tabular}
\label{table:manualEvaluation}
\vspace{-7mm}
\end{table}

We first evaluate the quality of our hate set by manually annotating a random sample of Gab and Twitter posts. We take 100 random posts that were marked as hate/non-hate by the framework described in Section~\ref{subsec:Hate}. Two authors of this work independently annotated these posts by hand as hate or non-hate according to their perspectives. We hide the previously assigned labels from the annotators. Table~\ref{table:manualEvaluation} shows the results of our manual evaluation for the given samples. 
We notice the occurrence of few false positive ($\leq 4\%$) and false negative ($\leq 5\%$) considering the agreement of the annotators. 

The annotators have agreed on the label of 383 posts, resulting in a Cohen's kappa coefficient $\kappa=0.92$. Assuming the labels created by the annotators of these 383 posts as the correct ones, 
the method to identify hate presented in this work was able to correctly classify 371 labels: accuracy 96.87\%, precision 96.32\%, and recall 97.34\%. These results reinforce the quality of the method to identify hate posts with minimal noise. 

\subsection{Types of Hate}

We associate one hate post with different types of hate, according to the types shown in Table~\ref{table:hateWords}: \textit{ethnicity}, \textit{class}, \textit{disability}, \textit{nationality}, \textit{religion}, \textit{gender}, and \textit{sexual orientation}. One hate post can be associated with up to seven types of hate (or none). Being associated with one type of hate does not necessarily imply that one post is being hateful towards that particular group, but rather it makes use of terms associated with that type of hate to write a comment that is perceived as rude, disrespectful, or unreasonable by different groups. 

\subsubsection{Frequent types of hate} 
\begin{figure}[t]
    \center {\includegraphics[scale=0.24]{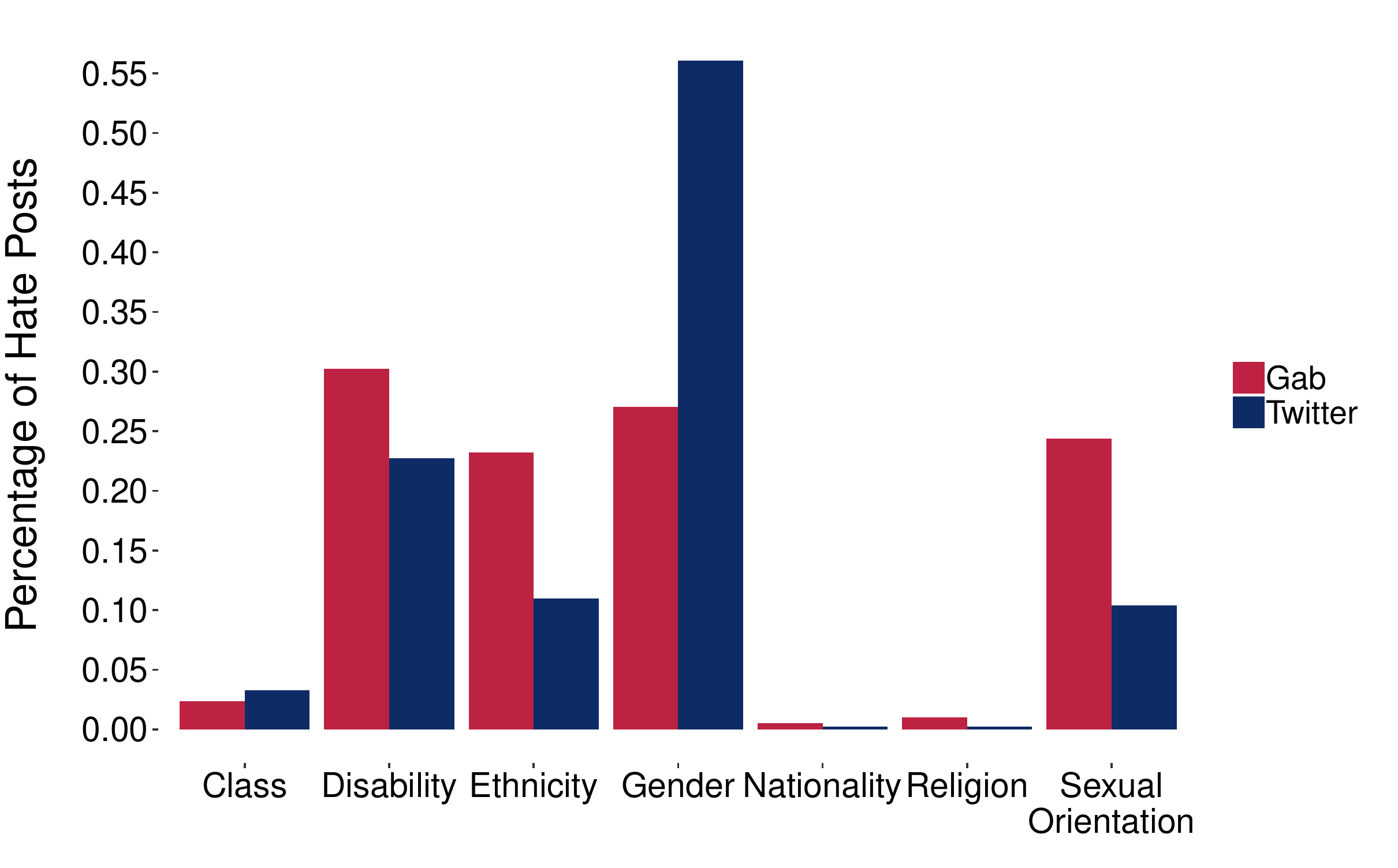}}
    \caption{Percentage of posts associated with each type of hate.}
    \label{fig:typesHate}
    \vspace{-5mm}
\end{figure}

Considering our hate set, Figure~\ref{fig:typesHate} shows the percentage of Gab and Twitter posts associated with each type of hate. Our findings show that Gab and Twitter posts are predominantly associated with disability and gender types of hate. It is interesting to notice that 56.06\% of hate tweets are associated with gender, followed by 22.74\% of hate tweets being related to disability, a difference of over 30\%. 

For Gab, this difference is smaller, as disability is associated with 30.25\% and is followed by gender, with 27.04\%. Gab has also a large number of hate posts associated with sexual orientation and ethnicity (over 20\% each). These results suggest that an environment which lacks moderation of content like Gab is more prone to the dissemination of hate speech of many types than a moderated one, which still needs to improve their hate speech policies and methods in order to avoid hate speech towards specific groups of people.

\subsubsection{Frequent hate terms} Table~\ref{table:frequentHateWords} shows the 5 most frequent hate terms in Gab and Twitter hate sets. The complete rankings have significant Kendall rank correlation coefficient (0.76). We observe that all gender related hate terms appear proportionally more on Twitter than Gab, which helps explaining the larger number of gender related hate posts observed in Twitter in comparison with Gab on Figure~\ref{fig:typesHate}. The term \textit{c*nt} appears in more than 45\% of Twitter hate posts whereas this number for Gab drops to near 22\%. \textit{T*at} and \textit{d*ke} are the others gender related terms on the top 10 which appear proportionality more on Twitter hate posts than Gab's, corroborating our findings on the analysis of linguistic differences between these networks.

\begin{table}[t]
\centering
\caption{Top 5 most frequent hate terms in Gab and Twitter hate posts.}
\resizebox{\columnwidth}{!}{
\begin{tabular}{@{}llll@{}}
\toprule
\multicolumn{2}{c}{\textbf{Gab}}                                                                 &           \multicolumn{2}{c}{\textbf{Twitter}}                                                             \\ \midrule
\multicolumn{1}{c}{\textbf{term (\%)}} & \textbf{category}          &    \multicolumn{1}{c}{\textbf{term (\%)}}         & \textbf{category} \\
retarded (30.25)                         & disability                 &  c*nt (45.31)                              & gender                                                \\
fa**ot (23.96)                            & sexual orientation         &  retarded (22.74)                         & disability                                           \\
c*nt (22.10)                              & gender                     &  t*at (9.65)                              & gender                                             \\
ni**er (21.3)                            & ethnicity                  &  fa**ot (9.53)                           & sexual orientation                                    \\
t*at (4.68)                              & gender                     &  ni**er (8.52)                            & ethnicity                                             \\
\bottomrule
\end{tabular}}
\label{table:frequentHateWords}
\vspace{-6mm}
\end{table}

\subsection{Potential Limitations}
Our Twitter dataset is shaped by the limitations of getting a sample from all Twitter with the Streaming API~\cite{morstatter2013sample}. Moreover, hate speech classification is inherently difficult, as there is no universal definition for it and many important variables, such as context, are not easily measured. Our method to detect hate relies, as a first step, on an external API which might also lead to inaccurate toxicity scores for social media posts. Hosseini et al.~\cite{hosseini2017deceiving} have shown that subtle changes on highly toxic sentences may assign significantly lower scores to them, which may indicate that many posts could not be classified as hateful in our work. Nevertheless, we have showed that by building on previous our approach can accurately identify many forms of hate speech. 

\section{Conclusion}
\label{sec:7_Conclusion}

In this work, we provide an in-depth quantitative analysis of moderated data from Twitter and unmoderated data from Gab, a social network that received several criticisms regarding the content shared on it. We perform linguistic analysis and  conduct an investigation of hateful posts and the various shades of hate that are displayed in this right-leaning echo chamber. Our analysis on Gab, put into perspective with Twitter, showed that the unmoderated posts on Gab present more negative sentiment, higher toxicity, and different psycholinguistic features. Our findings support that unmoderated environments may have proportionally more hate speech. Furthermore, we categorize hate speech and its different forms in both environments, unveiling a highly toxic discourse in Gab in the many forms that hate speech can manifest itself. 

Our work makes an important step towards the development of automated hate speech detectors. We believe that an unmoderated hate dataset\footnote{\url{https://github.com/lhenriquecl/unmoderated-hate-dataset}}, as the one analyzed, can help the development of hate speech detection approaches in future works. For this reason, our final contribution consists of making our hate-labeled Gab posts available to the community. 

\section*{Acknowledgments}
This work was partially supported by the MPMG, project Analytical Capabilities, CNPq, CAPES, and Fapemig.

\bibliographystyle{IEEEtran}
\small{\bibliography{Bibliography-File}}

\end{document}